\documentclass[referee]{aa}

\usepackage{graphicx}
\usepackage{natbib}
\usepackage{latexsym}
\usepackage{amsmath}
\usepackage{amssymb}
\usepackage{amstext}
\usepackage{color}
\usepackage{psfrag}
\usepackage{txfonts}

\def\k{km s$^{-1}$}
\def\ks{km s$^{-1}$~}
\def\d{$^\circ$}
\def\m{$^\prime$}
\def\s{$^{\prime\prime}$}
\def\hh{$^{\mathrm h}$}
\def\mm{$^{\mathrm m}$}
\def\ss{$^{\mathrm s}$}
\def\cm3{cm$^{-3}$}

\def\2{$^{12}$CO}
\def\3{$^{13}$CO}
\def\msol{M$_\odot$}

\def\cm2{cm$^{-2}$}
\begin{document}

\title{The molecular clump towards the eastern border of SNR G18.8+0.3}
\author {S. Paron \inst{1,2,3}
\and M. E. Ortega \inst{1}
\and A. Petriella \inst{1,3}
\and M. Rubio \inst{4}
\and G. Dubner \inst{1}
\and E. Giacani \inst{1,2}
}

\institute{Instituto de Astronom\'\i a y F\'\i sica del Espacio (IAFE),
             CC 67, Suc. 28, 1428 Buenos Aires, Argentina\\
             \email{sparon@iafe.uba.ar}
\and FADU - Universidad de Buenos Aires, Ciudad Universitaria, Buenos Aires
\and CBC - Universidad de Buenos Aires, Ciudad Universitaria, Buenos Aires
\and Departamento de Astronom\'{\i}a, Universidad de Chile, Casilla 36-D, Santiago, Chile}

\offprints{S. Paron}

   \date{Received <date>; Accepted <date>}

\abstract{}{The eastern border of the SNR G18.8+0.3, close to an HII regions complex, 
is a very interesting region to study the molecular gas that it is probably in 
contact with the SNR shock front.}{We observed the aforementioned region using the Atacama 
Submillimeter Telescope Experiment (ASTE) in the \2 J=3--2, \3 J=3--2,
HCO$^{+}$ J=4--3, and CS J=7--6 lines with an angular resolution of 22\s. 
To complement these observations, we analyzed IR, submillimeter and radio continuum archival data.}
{In this work, we clearly show that the radio continuum ``protrusion'' that was early 
thought to belong to the SNR is an HII regions complex deeply embedded in a molecular
clump. The new molecular observations reveal that this dense clump, belonging to an extended 
molecular cloud that surrounds the SNR southeast border, is not physically
in contact with SNR G18.8+0.3, suggesting that the SNR shock front have not yet reached it or maybe
they are located at different distances. 
We found some young stellar objects embedded in the molecular clump, suggesting that their 
formation should be approximately coeval with the SN explosion. 
%The proposed scenario is that the SNR G18.8+0.3 was the first star exploding in an active region of massive stars.
}{}

\titlerunning{SNR G18.8+0.3 and the eastern molecular clump}
\authorrunning{S. Paron et al.}


\maketitle

\section{Introduction}

The conversion of gas into stars involves a diversity of objects 
(molecular clouds, dust, magnetic fields, etc.) and several highly nonlinear and 
multidimensional  dynamical processes (turbulence, self-gravity, etc.) operating at different 
scales, which are still far from understood in spite 
of all the theoretical and observational advances. Detailed multiwavelength studies at 
different spatial scales, from large clouds to star embryos, 
are very helpful to understand precisely how the gas and dust coalesces until forming new stars. 
Particularly, the investigation of the properties 
of the medium from which stars form, is a useful way to know the initial conditions that  
favour star formation and the most favourable mechanisms 
to trigger the process. 

In this context, the molecular cloud suggested to be interacting with the SNR G18.8+0.3 
\citep{dubner99, dubner04, tian07}, which harbours IRAS pointlike 
sources compatible with the characteristics of protostellar candidates (IRAS sources 
whose colors correspond to those of ultracompact HII regions, i.e. sites of 
recent massive star formation, \citet {wood89}) located very close to the SNR shock, is 
an interesting target. Its study allows us not only to analyze the properties 
of the interstellar matter around pre-stars, but also to explore if there exist some 
relation between the new born stars and the SNR. Studies of the molecular
gas in regions suspected to be stellar nurseries, performed with intermediate spatial 
resolution, are especially useful to pinpoint the best candidates to 
pursue the millimetric and submillimetric studies with the unprecedented resources  
provided by ALMA.

The SNR G18.8+0.3 has a peculiar morphology with the eastern and southern flanks strongly 
flattened, while it fades to the west (Figure 1), 
suggesting a marked density gradient in the  ambient gas. Based on VLA radio 
continuum observations, \citet{dubner96} pointed out the existence of a ``protrusion'' 
emerging from the eastern border of the SNR, near  18\hh 24\mm 14\ss, -12\d 28\m 30\s (J2000). 
Subsequent observations performed with better angular resolution resolved, 
as described below, this region into several HII regions. In this paper, we analyze the 
molecular emission in this spot.

\begin{figure}[h]
\centering
\includegraphics[width=8cm]{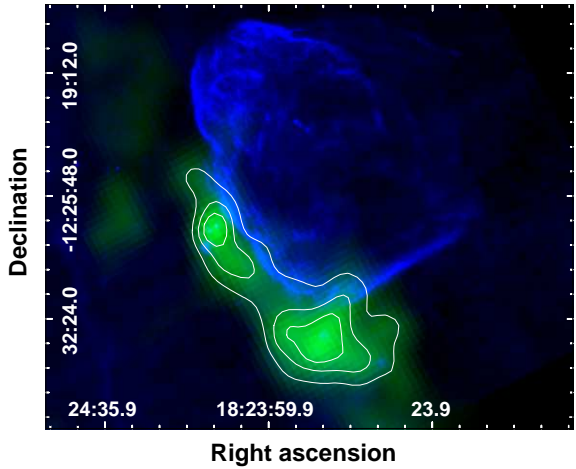}
\caption{The SNR G18.8+0.3 radio continuum emission at 20 cm is displayed in blue and the \3 J=1--0 
emission, extracted from the GRS \citep{simon01},
which was averaged between 17 and 22 \ks is presented in green with withe contours. }
\label{13coBig}
\end{figure}

\citet{arnal93} early reported the detection of \2 line emission along all the eastern border 
of the SNR G18.8+0.3, suggesting that the cloud might have been shocked 
by the expanding remnant. Later, \citet{dubner99} carried out observations in HI, \2 and \3 
in the direction to G18.8+0.3, concluding that the SNR explosion took place 
near the border of a pre-existing cloud, driving a slow shock (v $\sim$ 10 \ks) into the cloud. 
Based on the systemic velocity v$_{\rm{LSR}} \sim 19$ \ks of the interacting 
molecular cloud, the distances of 1.9 and 14.1 kpc were estimated by applying a Galactic 
circular rotation model. In that paper, the distance ambiguity was wrongly  
interpreted favouring the closest option, a problem  solved in \citet{dubner04}, where the 
far distance of 14 kpc was confirmed for G18.8+0.3. More recently, based on \3 
emission and HI absorption techniques, \citet{tian07} established 6.9 kpc and 15 kpc as 
the lower and upper limits for the distance to this SNR. 
%By assuming a distance 
%of 14 kpc, the dynamical age of the SNR G18.8+0.3 turns out to be about 10$^5$ yrs. This is an important parameter to compare with the contraction timescales of the young 
%stellar objects and decide whether the passage of a SN shock might have been responsible for the formation of new stars. 

\section{Presentation of the investigated region}
\label{present}

\begin{figure}[h]
\centering
\includegraphics[width=11cm]{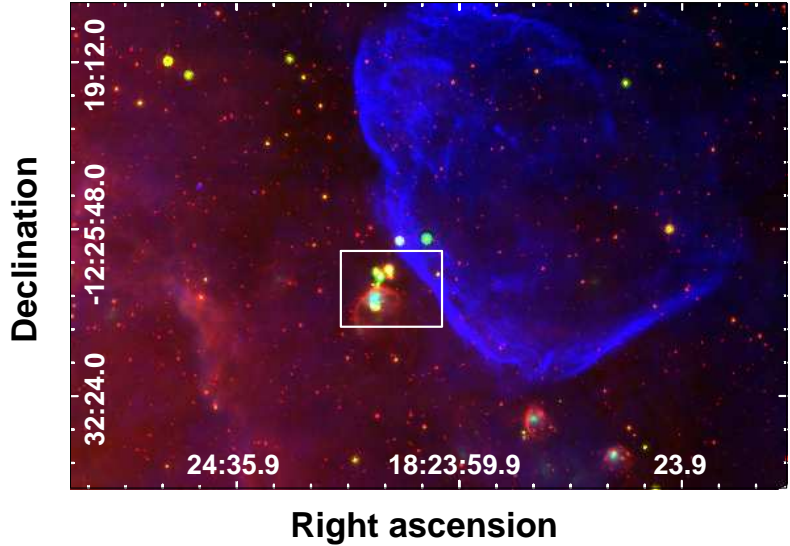}
\caption{Three-color image of the SNR G18.8+0.3 and its surroundings. Again the radio 
continuum emission at 20 cm is presented in blue,
the infrared bands at 8 $\mu$m and 24 $\mu$m are shown in red and green, respectively. 
The white rectangle shows the studied region.}
\label{presBig}
\end{figure}

\begin{figure}[h]
\centering
\includegraphics[width=9cm]{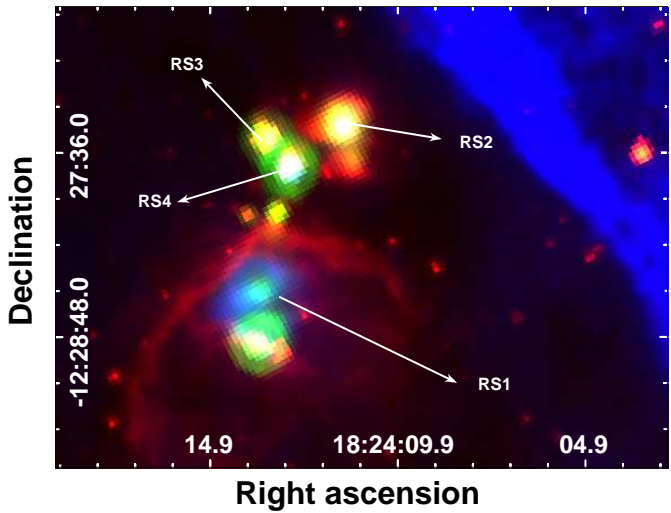}
\caption{Three color image of the region marked with the white box in Fig. \ref{presBig}. Radio continuum at 20 cm = blue,
 8 $\mu$m = red, and 24 $\mu$m = green. The catalogued radio sources described
in the text are marked.}
\label{presSmall}
\end{figure}

Figure \ref{13coBig} shows the SNR G18.8+0.3 as seen in the radio continuum emission at 20 cm extracted from
the MAGPIS \citep{helfand06} (in blue) and its molecular environment in the \3 J=1--0 
emission averaged between 17 and 22 \ks (in green). The \3 data, with an angular resolution
of 46\s, were extracted from the Galactic Ring Survey (GRS; \citealt{simon01}). 
The image shows a clear large-scale morphological correspondence between the SNR southeastern border and the 
molecular gas, strongly suggesting an interaction between them, as was suggested in previous works.
Figure \ref{presBig} shows the SNR G18.8+0.3 radio continuum (again in blue) and its IR environment 
as seen in the {\it Spitzer}-IRAC band at 8 $\mu$m (red), 
and the {\it Spitzer}-MIPS band at 24 $\mu$m (green). 
The box indicates the region studied in this work. Figure \ref{presSmall} 
displays a zoom-in of this region showing the presence 
of several catalogued radio sources. The source RS1 is catalogued in the HII Region Discovery 
Survey \citep{anderson11} as an irregular bubble (G18.751+0.254) with a 
recombination line at the LSR velocity 19.1 \k, thus confirming that it is immersed in the same
molecular cloud adjacent to the SNR G18.8+0.3. 
In Fig. \ref{presSmall}, it can be appreciated the radio continuum emission of this HII region 
surrounded by 8 $\mu$m IR emission with a 
bright center emitting in 24 $\mu$m, as is usually observed in infrared dust bubbles 
\citep{church06,church07}. The sources RS2, RS3, and RS4, appear as discrete radio sources 
in the MAGPIS 20 cm Survey. In particular the source RS4, catalogued as a radio compact HII 
region \citep{giveon05}, lies
at the same position as the 870 $\mu$m continuum source G18.76+0.26. This source, 
observed with the Large APEX Bolometer Camera (LABOCA) and catalogued 
in the ATLASGAL, has an angular extension of 59\s$\times$42\s~\citep{schuller09}. From 
observations of the NH$_{3}$ (1,1) line, these authors report a v$_{\rm LSR} = 20.8$ \ks for 
G18.76+0.26, concluding that this source is located at the distance of about 14 kpc.

We conclude that the molecular complex abutting the eastern border of the SNR G18.8+0.3 is a 
very rich region populated by several HII 
regions. The HI absorption analysis performed by \citet{tian07} towards this 
region (region 6 in their work), gives an spectrum with the same absorption features as 
the spectra towards several regions over the SNR, strongly suggesting that this HII regions 
complex is located at the same distance as the SNR. 
%Nowadays it is well known that massive stars usually form in clusters and/or stellar associations (e.g. \citealt{mckee07}), hence it is expectable that several HII regions in 
%different evolutionary stages and  
%also SNRs belong to a same Galactic neighborhood located at the same distance. This statement and the reported distance of about 14 kpc for the SNR G18.8+0.3 and the 870 $\mu$m 
%continuum source G18.76+0.26 which is related to RS4 strongly suggest that all sources shown in Figs. \ref{presBig} and \ref{presSmall} 
%are not only close in the plane of the sky, but in physical vicinity. 
Thus, we adopt a distance of about $14\pm1$ kpc for the SNR, the molecular cloud and the HII 
regions complex. The error bar of 2 kpc comes by considering non-circular motions in the Galactic 
rotation model towards this region of the Galaxy.

\section{Observations}

The molecular observations presented in this work were performed on June 12 and 13, 2011 
with the 10 m Atacama Submillimeter
Telescope Experiment (ASTE; \citealt{ezawa04}). We used the CATS345 GHz band receiver, which is a two-single
band SIS receiver remotely tunable in the LO frequency range of 324-372 GHz. We simultaneously
observed \2 J=3--2 at 345.796 GHz and HCO$^{+}$~J=4--3 at
356.734 GHz, mapping a region of 240\s~$\times$ 150\s~centered at RA $=$ 18\hh 24\mm 10.9\ss, 
dec. $= -$12\d 28\m 22.0\s, J2000.
We also observed \3 J=3--2 at 330.588 GHz and CS J=7--6 at 342.883 GHz mapping a region 
of 120\s~$\times$ 120\s~centered at 
RA $=$ 18\hh 24\mm 10.9\ss, dec. $= -$12\d 27\m 20.0\s. The mapping grid spacing was 20\s~in 
all cases and the integration time was 30 and 60 sec per pointing in each case, 
respectively. All the observations were performed in position switching mode. 

We used the XF digital spectrometer with a bandwidth and spectral resolution set to 
128 MHz and 125 kHz, respectively.
The velocity resolution was 0.11 \ks and the half-power beamwidth (HPBW) was 22\s~at 345 GHz. 
The weather conditions were optimal and the system temperature
varied from T$_{\rm sys} = 150$ to 200 K. The main beam efficiency was $\eta_{\rm mb} \sim 0.65$. 
All quoted numbers for the line temperatures along this work were 
corrected for the antenna efficiency, i.e. in all cases they are the main brightness temperature.
The spectra were Hanning smoothed to improve the signal-to-noise ratio and only linear or/and some third order
polynomia were used for baseline fitting.
The data were reduced with NEWSTAR\footnote{Reduction software based on AIPS developed at NRAO, 
extended to treat single dish data
with a graphical user interface (GUI).} and the spectra processed using the XSpec software 
package\footnote{XSpec is a spectral line reduction package for astronomy which has been 
developed by Per Bergman at Onsala Space Observatory.}.

\begin{figure}[h]
\centering
\includegraphics[width=5cm,angle=-90]{fig12CO2020.ps}
\caption{\2 J=3--2 spectrum obtained towards the offset ($+$20\s,$+$20\s) position relative to
RA $=$ 18\hh 24\mm 10.9\ss, dec. $= -$12\d 28\m 22.0\s, J2000.}
\label{espec12co}
\end{figure}

\begin{figure}[h]
\centering
\includegraphics[width=10cm]{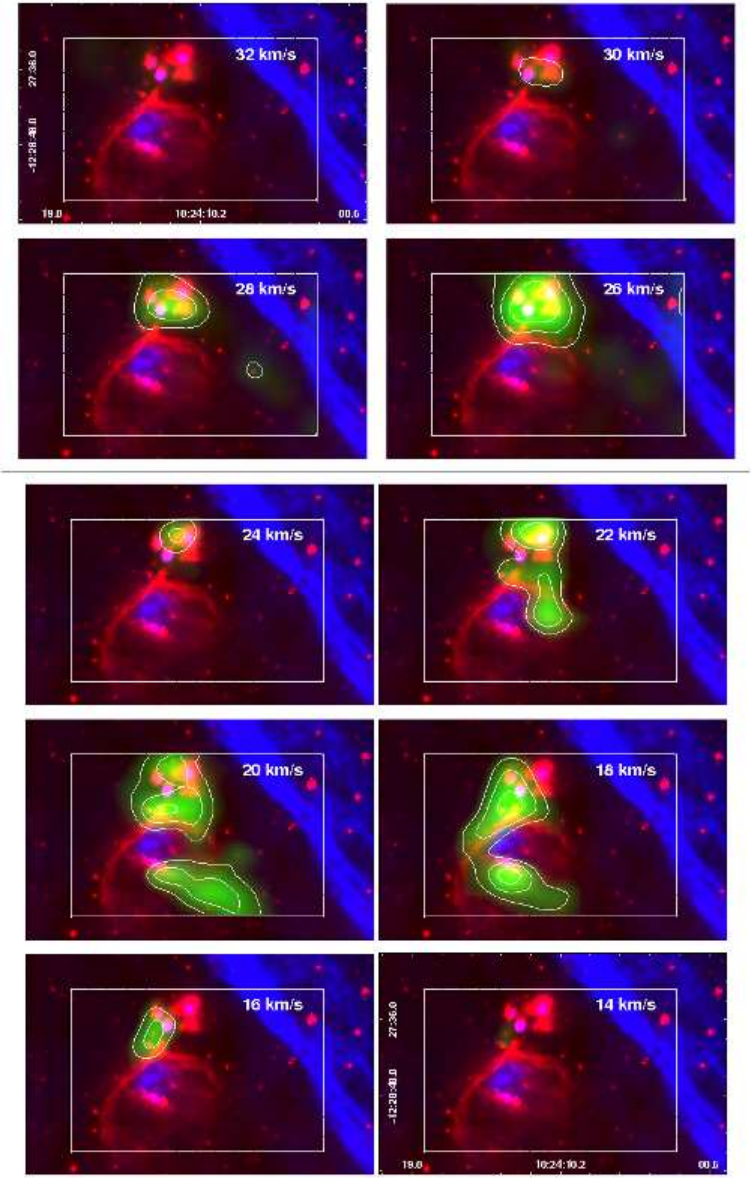}
\caption{Integrated velocity channel maps of the \2 J=3--2 emission every 2 \ks (in green). As previous figures, in red is shown
the IRAC emission at 8 $\mu$m and in blue the radio continuum emission at 20 cm. The contours levels in the upper panels (from
32 to 26 \k) are 1.8, 3.6 and 7.0 K \k, while the contours levels in the bottom panels (from 24 to 14 \k) are 11.5, 15.0, and 19.5 K \k.
The dashed rectangle shows the mapped region.
}
\label{panels}
\end{figure}

\section{Molecular gas}
\label{mol}

\begin{figure}[h]
\centering
\includegraphics[width=10cm]{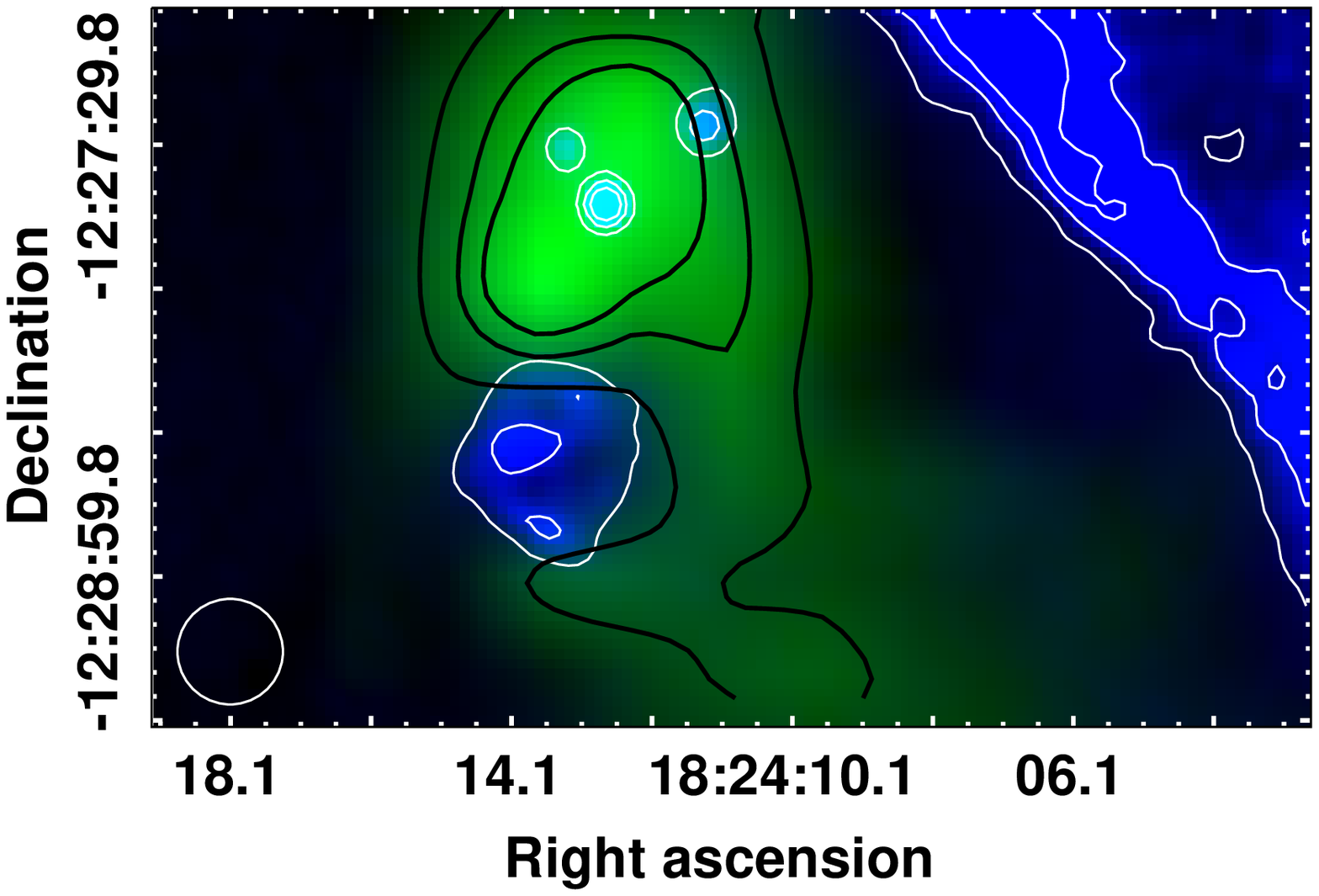}
\caption{The \2 J=3--2 emission integrated between 10 and 30 \ks is presented in green with black 
contours with levels of 46, 65, and 80 K \k. The radio continuum emission at
20 cm is shown in blue with white contours of 2.3, 6, and 11 mJy beam$^{-1}$. The angular resolutions 
are about 20\s~and 6\s~for the molecular and radio continuum emissions, 
respectively. The beam of the molecular observations is included towards the bottom left corner.} 
\label{12cointeg}
\end{figure}

Figure \ref{espec12co} displays a typical \2 J=3--2 spectrum obtained towards the surveyed region. 
This spectrum, corresponding to the offset position ($+$20\s,$+$20\s), exhibits 
two components: a weak one at $\sim$5 \k, and the main component centered at $\sim19$ \k. The weak 
component corresponds to local gas emission and will not be further 
considered. The main 
component goes from $\sim$10 to 30 \k, and represents the molecular gas that it is very 
likely related to the SNR and the HII regions complex.  Figure \ref{panels} shows the gas distribution (in green)
as seen in the \2 J=3--2 emission integrated every 2 \ks from 14 to 32 \k.
After integrating the \2 J=3--2 
emission from 10 to 30 \ks we obtain the molecular clump shown in Fig. \ref{12cointeg}. 
The HII region G18.751+0.254 (source RS1 in Fig. \ref{presSmall}, seen in blue in 
Fig. \ref{espec12co}) appears surrounded by the molecular emission, whose peak coincides with 
the region where the sources RS2, RS3 and RS4 are located. 
Taking into account that the southeastern border of SNR G18.8+0.3 is surrounded by an extended 
molecular cloud detected in lower molecular transitions (see Fig. \ref{13coBig}), 
we conclude that with the present observations we are analyzing a molecular clump belonging to 
this extended cloud.

\begin{figure}[h]
\centering
\includegraphics[width=10cm]{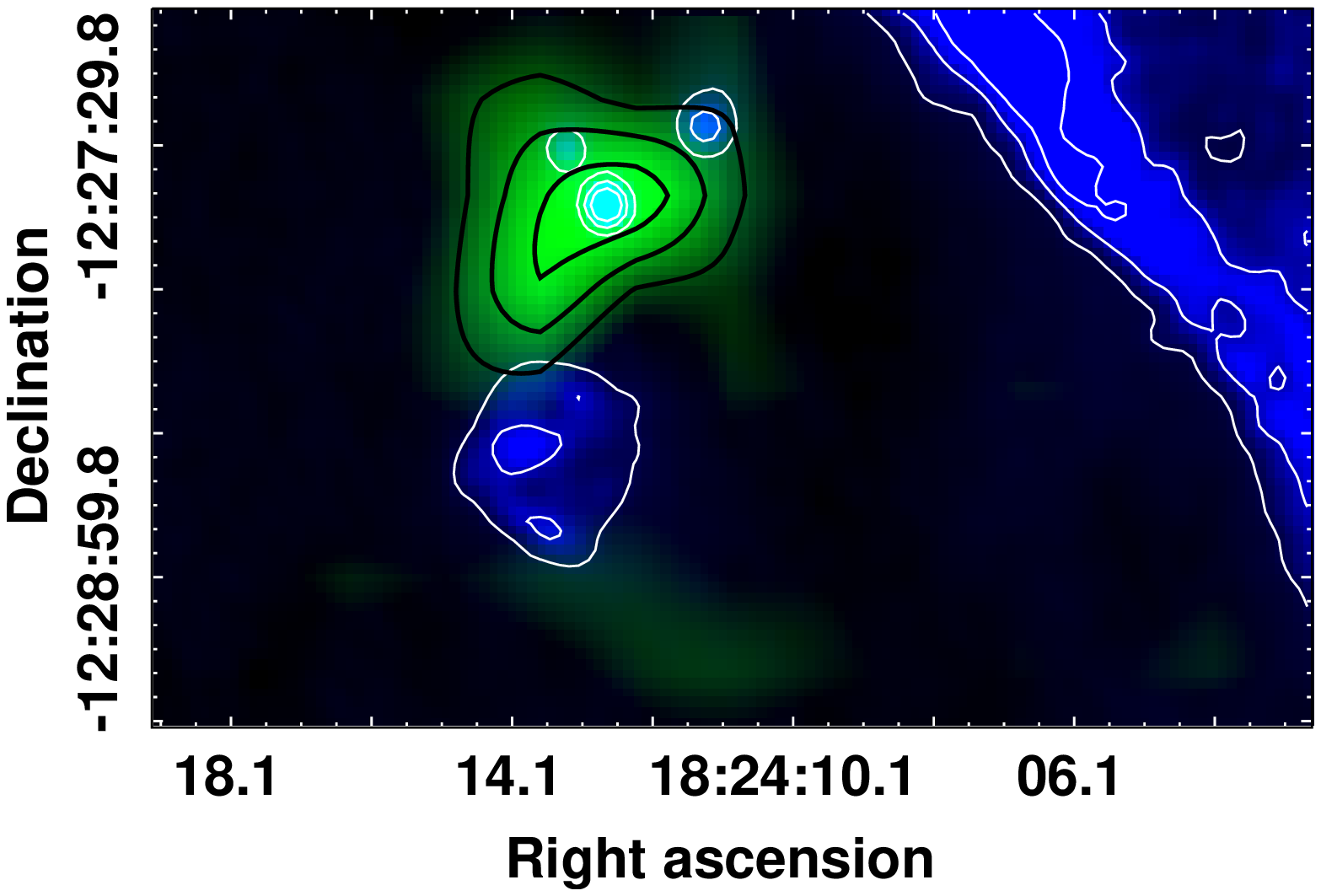}
\caption{The HCO$^{+}$ J=4--3 emission integrated between 10 and 30 \ks is displayed in green 
with black contours with levels of 4, 6, and 8 K \k. The radio continuum emission at
20 cm is shown in blue with white contours of 2.3, 6, and 11 mJy beam$^{-1}$. }
\label{hco+}
\end{figure}

\begin{figure}[h]
\centering
\includegraphics[width=10cm]{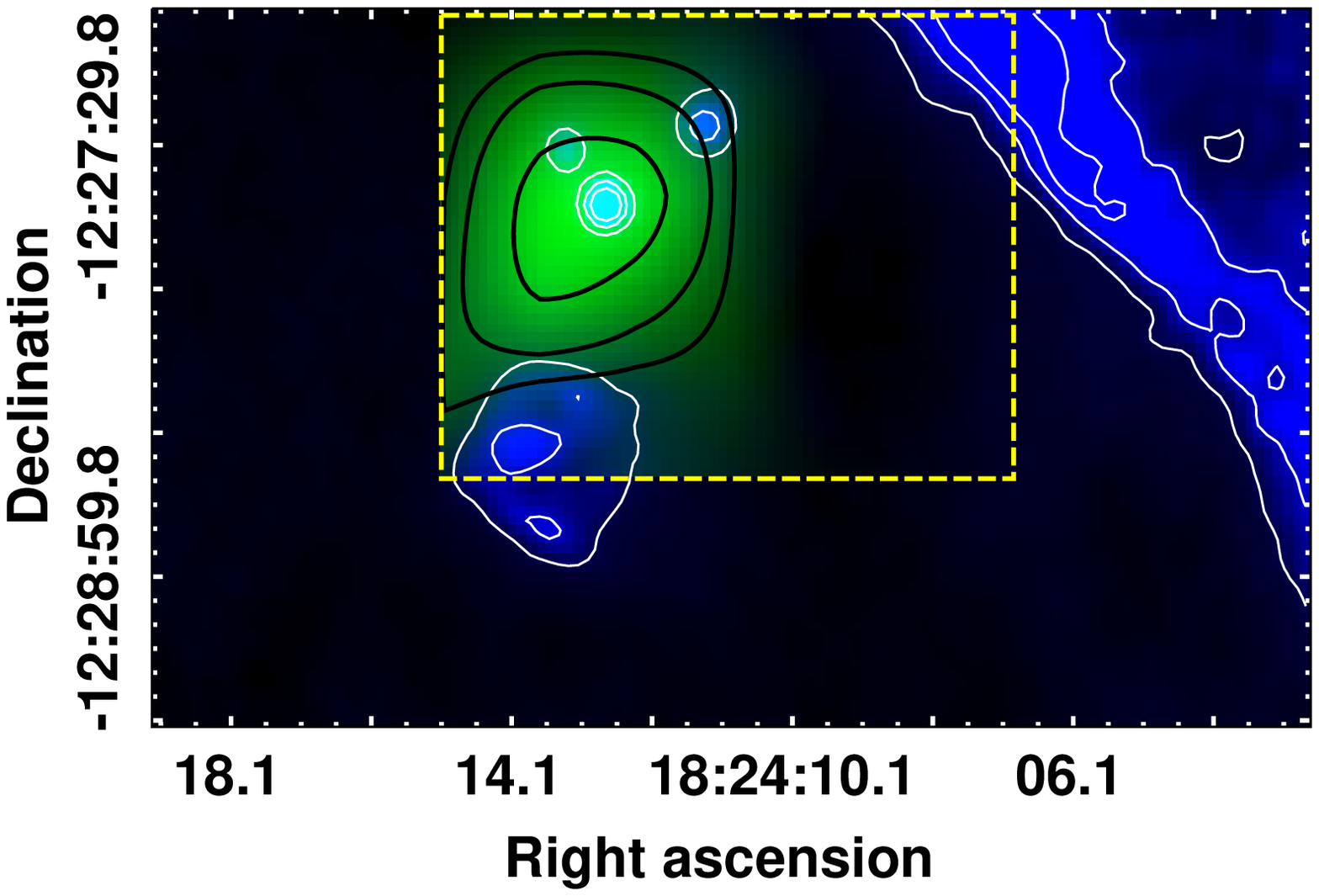}
\caption{The \3 J=3--2 emission integrated between 10 and 30 \ks is presented in green with 
black contours with levels of 15, 20, and 30 K \k. The radio continuum emission at
20 cm is shown in blue with white contours of 2.3, 6, and 11 mJy beam$^{-1}$. The dashed yellow 
rectangle represents the region surveyed in this line. }
\label{13co}
\end{figure}

From the inspection of the other observed molecular transitions, we find that the HCO$^{+}$ J=4--3 
and \3 J=3--2 lines are only bright near the peak of the molecular clump, i.e. towards 
the densest region. Figures \ref{hco+} and \ref{13co} display the 
HCO$^{+}$ J=4--3 and \3 J=3--2 emissions, respectively, integrated between 10 and 30 \k. 
In Fig. \ref{13co}, the dashed yellow rectangle represents the region surveyed 
in the \3 line.

Based on the presented molecular maps we suggest that RS1 has shaped the surrounding gas, while
the other radio sources, likely younger HII regions, are still embedded in the densest portion of 
the molecular clump.  On the other hand it is important to note that the analyzed 
molecular clump is not in physical contact with the SNR G18.8+0.3 border, suggesting that the SNR 
shock front have not yet reached it or the clump is located at a different distance.

\begin{figure}[h]
\centering
\includegraphics[width=12cm]{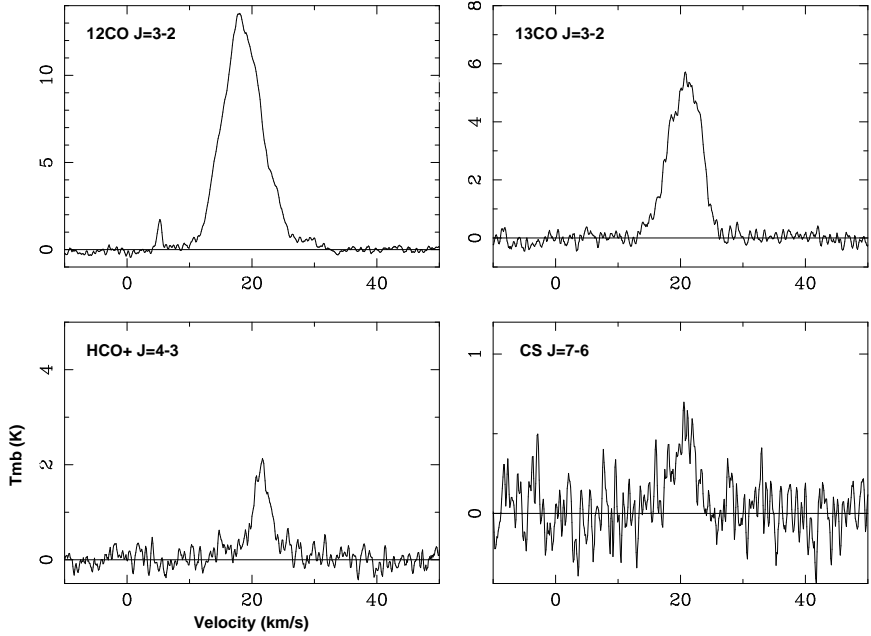}
\caption{Spectra of each molecular species obtained towards the molecular clump peak.  
The \2, \3, and HCO$^{+}$ lines have very high S/N ratio, while the CS emission has 
a S/N ratio of about 2.2.}
\label{spec}
\end{figure}

Figure \ref{spec} shows the spectra of each molecular species obtained towards the peak position of 
the molecular clump. The parameters determined from Gaussian 
fitting of these lines are presented in Table \ref{lines}.
T$_{\rm mb}$ represents the peak brightness temperature, V$_{\rm LSR}$ the central velocity referred 
to the Local Standard of Rest, and $\Delta$v is the FWHM line width.
Errors are formal 1$\sigma$ value for the model of the Gaussian line shape. Additionally in this 
table are included the integrated intensities (I). It is noticeable 
that the HCO$^{+}$ J=4--3 line appears significantly narrower than the CO lines, suggesting that 
the dense gas, mapped by the HCO$^{+}$, occupies a different volume than the gas 
mapped by the \2 and \3 lines. 
Regarding the CS J=7--6 line, the data present some hints of emission towards this region. However 
the poor S/N ratio achieved in the observations of this transition does not allow us
to perform a trustable analysis about this molecular species. The CS spectrum presented in 
Fig. \ref{spec}, which was obtained towards the peak of the molecular clump,  
has a S/N ratio of about 2.2. 

\begin{table}
\caption{Observed and derived parameters of the molecular lines shown in Figure \ref{spec}.}
\centering
\begin{tabular}{lcccc}
\hline
Emission & T$_{\rm mb}$ & V$_{\rm LSR}$ & $\Delta$v  &    I    \\
         &  (K)         & (\k)          & (\k)       &  (K \k) \\
\hline
$^{12}$CO J=3--2 & 12.9 $\pm$0.1   &  18.5 $\pm$0.5 & 7.4 $\pm$0.1 & 104.7 $\pm$1.7 \\
\hline
$^{13}$CO J=3--2  & 5.5 $\pm$0.2   &  20.7 $\pm$0.5 & 6.0 $\pm$0.2 &  35.0 $\pm$1.8  \\
\hline
HCO$^{+}$ J=4--3  & 1.8 $\pm$0.3   &  21.5 $\pm$0.7 & 3.6 $\pm$0.5 &  9.0 $\pm$1.6 \\
\hline
\end{tabular}
\label{lines}
\end{table}

\subsection{Column densities and abundances}

To estimate the molecular column densities and abundances towards the clump we assume local
thermodynamic equilibrium (LTE) and a beam filling factor of 1.
From the peak temperature ratio between the CO isotopes ($^{12}$T$_{mb}$/$^{13}$T$_{mb}$),
it is possible to estimate the optical depths from (e.g. \citealt{scoville86}):
\begin{equation}
\frac{^{12}{\rm T}_{mb}}{^{13}{\rm T}_{mb}} = \frac{1-exp(-\tau_{12})}{1-exp(-\tau_{12}/X)}
\label{eq1}
\end{equation}
where $\tau_{12}$ is the optical depth of the \2 gas and $X =$ [\2]/[\3] is the isotope abundance ratio. 
The [\2]/[\3] ratio can be estimated from the relation
[$^{12}$C]/[$^{13}$C] $ = 6.21 \times D_{G.C.} + 18.77 $ \citep{milam05}, where $D_{G.C.} = 6.9$ kpc
is the distance between the source and the Galactic Center, yielding
[$^{12}$C]/[$^{13}$C] $ = 61\pm14$.  According to \citet{milam05} and \citet{savage02},
the [$^{12}$C]/[$^{13}$C] isotope ratio exhibits a noticeable gradient with distance
from the Galactic center, i.e. is strong dependent with $D_{G.C.}$.
Then, from eq. \ref{eq1}, the \2 J=3--2 optical depth is $\tau_{12} \sim 30$, while the \3 J=3--2 optical depth is $\tau_{13} \sim 0.5$, 
revealing that the \2  line appears optically thick, while the \3 line is optically thin.
Thus, we calculate the excitation temperature from
\begin{equation}
T_{ex}(3 \rightarrow  2) = \frac{16.95 {\rm K}}{{\rm ln}[1 + 16.59 {\rm K} / (T_{\rm max}(^{12}{\rm CO}) + 0.036 {\rm K})]}
\label{eq2}
\end{equation}
obtaining $T_{ex} \sim 20$ K. Then, we derive the \2 column density from:
\begin{equation}
{\rm N(^{12}CO)} = 7.96 \times 10^{13}~e^{\frac{16.6}{T_{ex}}}\frac{T_{ex} + 0.92}{1 - exp(\frac{-16.6}{T_{ex}})} \int{\tau_{12}{\rm dv}}
\label{eq4}
\end{equation}
where, taking into account that $\tau \geq 1$ we use the approximation:
\begin{equation}
\int{\tau ~{\rm dv}} = \frac{1}{J(T_{ex}) - J(T_{\rm BG})} \frac{\tau}{1-e^{-\tau}} \int{{\rm T_{mb} ~dv}}
\label{eq5}
\end{equation}
with
\begin{equation}
J(T) = \frac{h\nu/k}{exp(\frac{h\nu}{kT}) - 1}
\label{eq6}
\end{equation}
Finally we obtain N(\2) $\sim 1.5 \times 10^{18}$ cm$^{-2}$.
In the case of the \3 J=3--2 line, we use:
\begin{equation}
{\rm N(^{13}CO)} = 8.28 \times 10^{13}~e^{\frac{15.87}{T_{ex}}}\frac{T_{ex} + 0.88}{1 - exp(\frac{-15.87}{T_{ex}})} \int{\tau_{13}{\rm dv}}
\label{eq3}
\end{equation}
and taking into account that this line appears optically thin, we use the approximation:
\begin{equation}
\int{\tau ~{\rm dv}} = \frac{1}{J(T_{ex}) - J(T_{\rm BG})} \int{{\rm T_{mb} ~dv}}
\label{eq9}
\end{equation}
yielding N(\3) $\sim 1.8 \times 10^{16}$ cm$^{-2}$.

The HCO$^{+}$ column density was derived from:
\begin{equation}
{\rm N(HCO^{+})} = 5.85 \times 10^{10}~e^{\frac{25.7}{T_{ex}}}\frac{T_{ex} + 0.71}{1 - exp(\frac{-17.12}{T_{ex}})} \int{\tau{\rm dv}}
\label{eq8}
\end{equation}
and by assuming that the HCO$^{+}$ J=4--3 is optically thin, we use the same approximation as used for the \3 line (eq. \ref{eq9}).
As excitation temperatures we use the range 20--50 K, obtaining  N(HCO$^{+}$) $\sim (3-5) \times 10^{12}$ cm$^{-2}$.

To estimate the molecular abundances it is necessary to have an H$_{2}$ column density value. As it will be
shown in detail in the next section (Sect. \ref{secc_bolo}), we independently estimate this parameter
through the millimeter continuum emission, obtaining N(H$_{2}$) $\sim 4.4\times10^{22}$~\cm2.
By using this value, we obtain the following molecular abundances:
X(\2) $\sim 3.4 \times 10^{-5}$, X(\3) $\sim 4.1 \times 10^{-7}$, and X(HCO$^{+}$) $\sim (0.7-1) \times 10^{-10}$.
The obtained HCO$^{+}$ abundance is very similar to the values derived by \citet{cortes10} and \citet{cortes11} towards
high-mass star-forming regions.

\section{Millimeter continuum emission}
\label{secc_bolo}

\begin{figure}[ht]
\centering
\includegraphics[width=11cm]{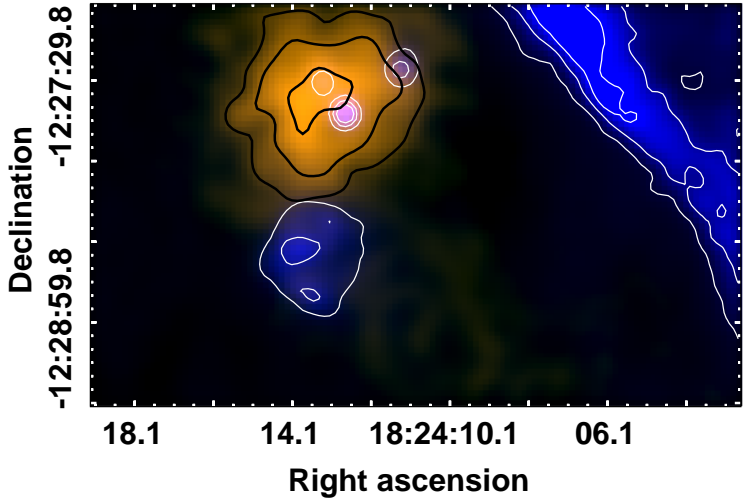}
\caption{The smoothed 1.1 mm dust continuum emission is presented in orange with black contours. 
The radio continuum emission at
20 cm is shown in blue with white contours of 2.3, 6, and 11 mJy beam$^{-1}$. }
\label{fig_bolo}
\end{figure}

We have also investigated the millimeter dust continuum emission at 1.1 mm using data from the Bolocam
Galactic Plane Survey. We found a source in positional coincidence with the \2 emission peak, BGPS
18.763+00.261 \citep{roso10}, which has a roughly elliptical morphology (41\s$\times$29\s). 
In Fig. \ref{fig_bolo},
we present a two-color image of the BGPS 1.1 mm emission and the radio continuum emission at 20 cm.
The positional agreement between the BGPS source and the molecular emission described above shows that 
the 1.1 mm emission originates in a densest portion of the molecular clump.

We estimate the mass of the molecular gas
associated with the BGPS source using the equations from
\citet{bally10}, M(H$_{2}) = 14.26D^{2}S_{1.1}(e^{13/T_{d}}-1)M_{\odot}$, 
where $D$ is the distance in kpc, $S_{1.1}$ is the flux density at 1.1 mm in Jy, and
$T_{d}$ is the dust temperature in K. Assuming that the dust and the gas are collisionally coupled, 
we can approximate 
$T_{d}=T_{K}$, where $T_{K}$ is the temperature of the gas. For $S_{1.1}$, we used the 80\s~aperture 
flux density, 
which seems appropriate for the dimensions of this particular BGPS source.
The flux value reported in \citet{roso10} was scaled with a calibration factor of
1.5 \citep{dunham11}. Thus, by adopting $D=14\pm1$~kpc, $T_{d}=20$~K, and $S_{1.1}=2.21$~Jy 
we obtain M(H$_{2}$)$ = 5700\pm800$~\msol. 
%For the number density, we approximate the volume of the clump with a sphere of
%radius equal to the deconvolved radius reported by \citet{roso10}, which is 35\s. We obtain
%n $\sim 1.8 \times 10^{3}$ cm$^{-3}$. 
In addition, we estimated the column density of the molecular gas by applying 
N(H$_{2}) = 2\times10^{22}S_{1.1}$~\cm2, obtaining N(H$_{2}) \sim 4.4\times10^{22}$~\cm2.

\section{The compact radio sources embedded in the molecular gas}
\label{sed}

\begin{figure}[h]
\centering
\includegraphics[width=10cm]{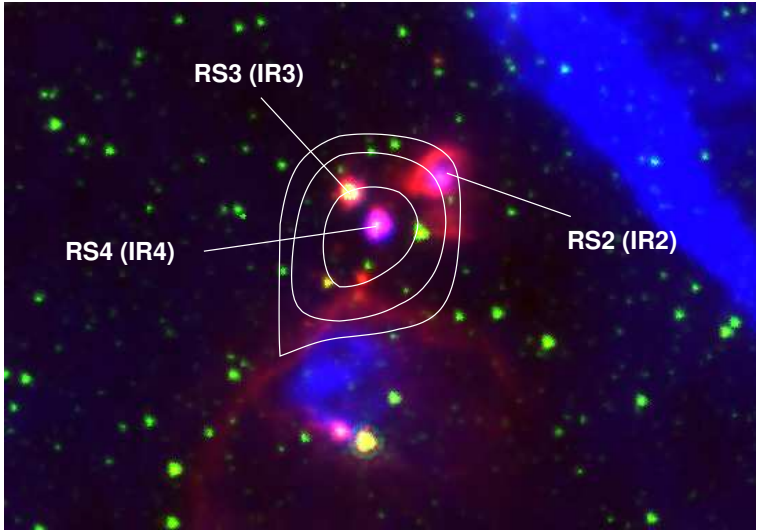}
\caption{Three-color image (blue = radio continuum at 20 cm; green = 4.5 $\mu$m; red = 8 $\mu$m). 
The radio sources RS2, RS3, and RS4 are 
indicated, and between brackets it is indicated the nomenclature of the associated infrared sources: 
WISE J182411.60-122726.0 (IR2), J182413.49-122730.0 (IR3), and J182412.89-122742.2 (IR4). The contours 
represent the $^{13}$CO J=3--2 emission integrated between 10 and 30~\k~with levels of 10, 15, and 30~K \k.}
\label{RS}
\end{figure}

Figure \ref{RS} shows a three-color image (blue = radio continuum at 20~cm; green = 4.5~$\mu$m; red
= 8~$\mu$m) of the studied region. The contours represent the
$^{13}$CO J=3--2 emission integrated from 10 to 30 \ks with levels of
15, 20, and 30 K \k. The radio sources, RS2, RS3, and RS4 (see Section
\ref{present}) are seen in projection onto the molecular gas
condensation. The brightest one in the radio band, RS4, positionally
coincides with the peak of the $^{13}$CO J=3--2 emission, while RS2 is
located towards the northwestern border of the clump. From the figure
it can be noticed that each radio source has 8~$\mu$m emission
associated with an excellent spatial correlation, which confirms the
thermal nature of the radio continuum emission. As noticed above, \citet{giveon05}
identified RS4 as a young compact HII region. In this work we identify two new likely young HII regions, RS2
(G018.765+0.262) and RS3 (G018.762+0.270). While in the case of RS3 the radio
continuum and the 8~$\mu$m emission have the same compact structure,
in RS2 the 8~$\mu$m emission exhibits a shell-like structure that
encircles the radio continuum emission.

From the WISE All-Sky Source Catalog \citep{wri10} we searched for infrared sources
related to radio ones. We identify the sources WISE
J182411.60-122726.0 (IR2), J182413.49-122730.0 (IR3), and J182412.89-122742.2 (IR4), which are 
related to the radio sources RS2, RS3, and RS4, respectively. 

\begin{figure}[h]
\centering
\includegraphics[width=6cm]{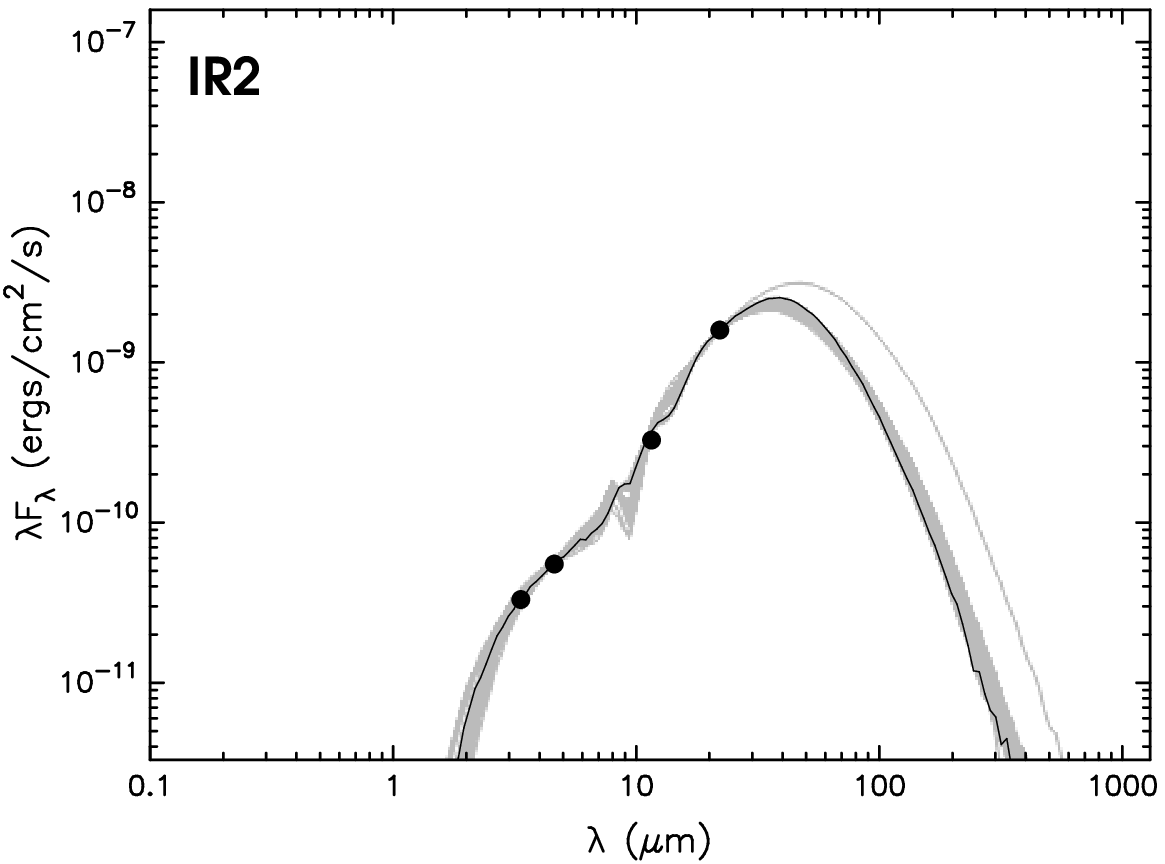}
\includegraphics[width=6cm]{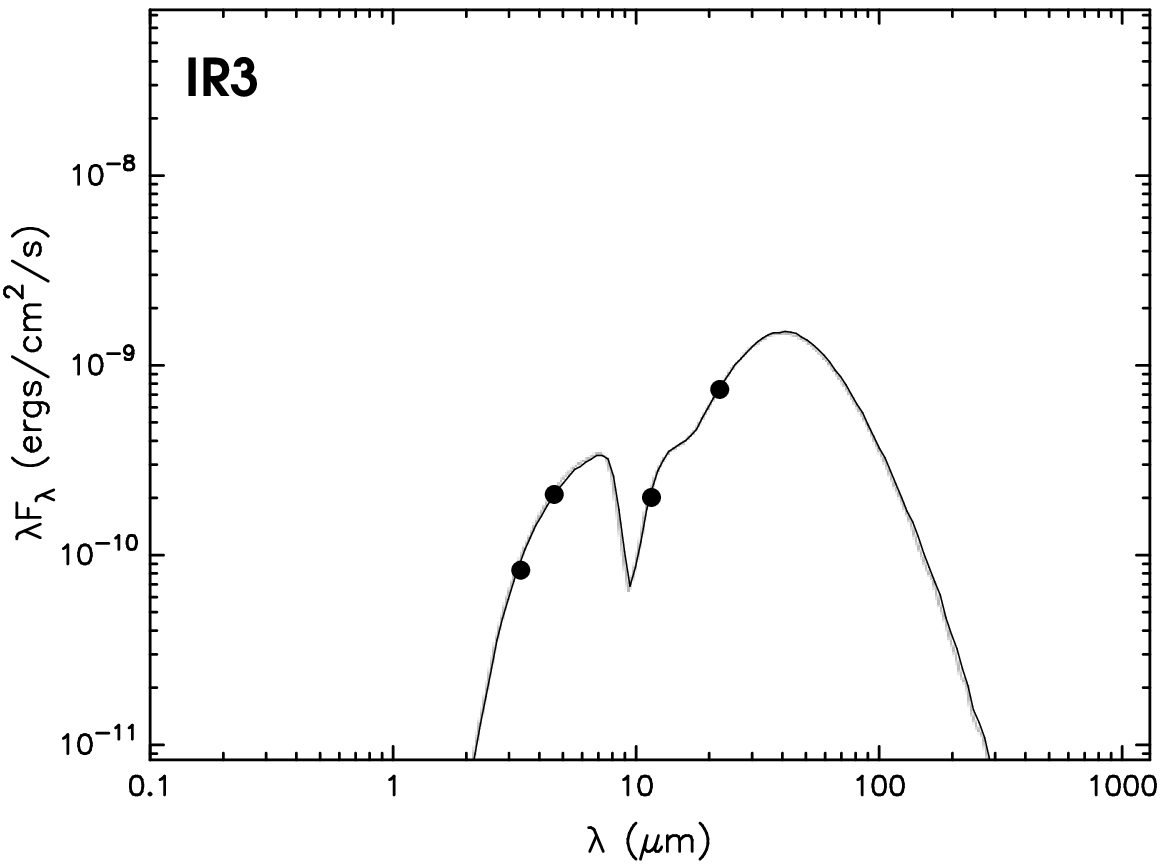}
\includegraphics[width=6cm]{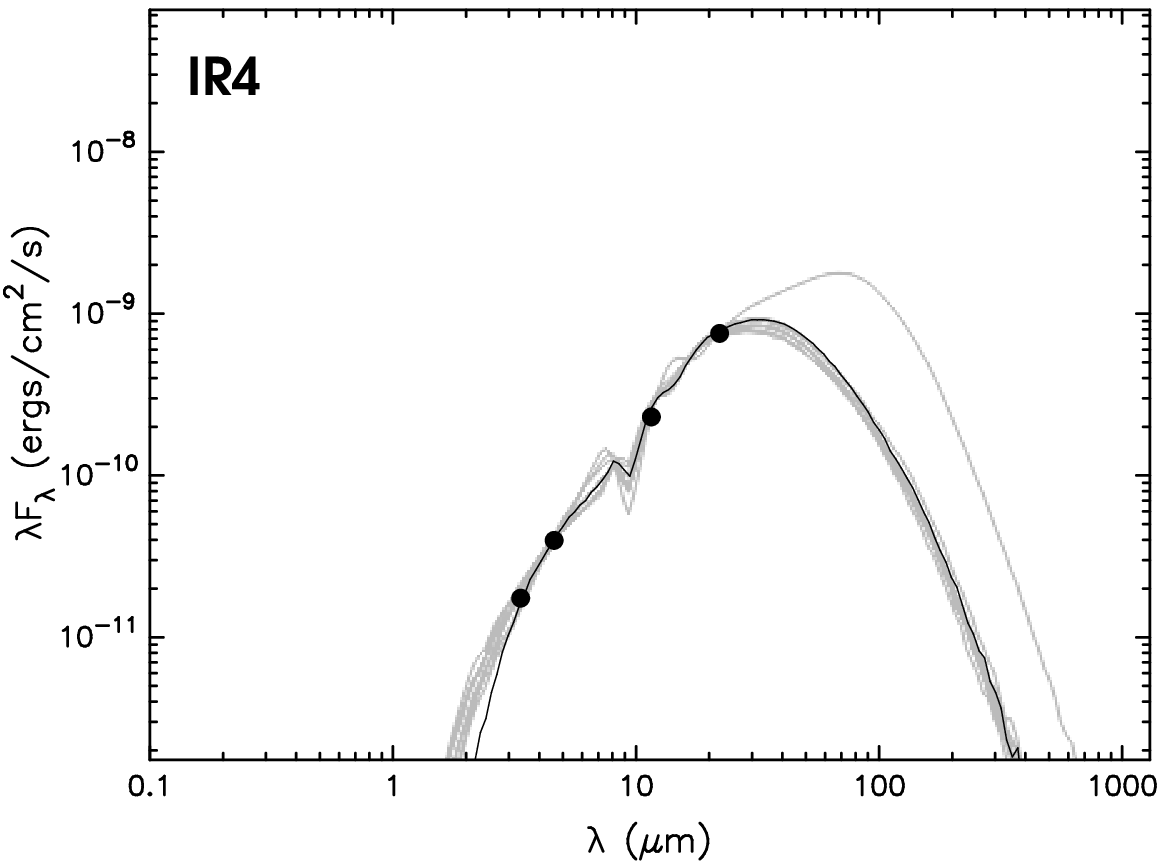}
\caption{SED models for IR2, IR3, and IR4. The circles
  indicate the measured fluxes. Black and gray solid curves represent
  the best-fit model and the subsequent good fittings models (with $\chi^2 - \chi^2_{best} < 3$),
  respectively.}
\label{SED}
\end{figure}

We performed a fitting of the spectral energy distribution
(SED) of IR2, IR3, and IR4, using the tool developed by
\citet{rob07}\footnote{http://caravan.astro.wisc.edu/protostars/} and adopting
an interstellar extinction in the line of sight, $A_v$, between
14 and 50 magnitudes. The lower limit of $A_v$ was chosen by
considering the typical value of one magnitude of absorption per
kpc. The upper limit corresponds to that $A_v$ derived from the molecular material,
estimated by using the relation $A_v = \frac{N(H_{2})}{0.94\times10^{21}}$ mag \citep{frer82}, 
with the N(H$_{2}$) derived in Sect. \ref{secc_bolo}. 
In Figure \ref{SED} we show the SEDs for the three sources
with the best-fitting model for each source (black curve) , and the subsequent good fitting models (grays curves)
with $\chi^2 - \chi^2_{best} < 3$ (where $\chi^2_{best}$ is the $\chi^2$ per
data point of the best-fitting model for each source).
To construct the
SED we consider the fluxes at the WISE 3.4, 4.6, 12, and 22~$\mu$m
bands. Table \ref{SEDpar} presents the main physical parameters of
each source as obtained from the best fitting model and the range from the subsequent models: star mass (Col. 2), star age (Col. 3), envelope accretion rate (Col. 4), 
bolometric luminosity (Col. 5),  $\chi^2$ per data point of the best fitting model (Col. 6), and the number of subsequent good 
fitting models (Col. 7).

\begin{table}
\scriptsize
\caption{Main physical parameters of IR2, IR3, and IR4 derived from the best fitting models for the SED of each source.} 
\centering
\begin{tabular}{ccccccccccc}
\hline
Source
&\multicolumn{2}{c}{$M_{\star}$}&\multicolumn{2}{c}{Age}&\multicolumn{2}{c}{$\dot
M_{env}$}&\multicolumn{2}{c}{$L$}& $\chi^2_{best}$& \#models\\
       &\multicolumn{2}{c}{[M$_{\odot}$]}&\multicolumn{2}{c}{[$\times
10^5$~yr]}&\multicolumn{2}{c}{ [$\times10^{-5}$ M$_{\odot}$yr$^{-1}$]}&\multicolumn{2}{c}{[$\times
10^{4} L_{\odot}$]} & & $\chi^2-\chi^2_{best}\leq 3$\\
\hline
     &  Best & Range & Best & Range & Best & Range & Best & Range & & \\
IR2 & 19 & 18-21 & 7 & 1-10  & 0  & 0-0.1   & 4 & 2-5 & 0.1 & 42  \\
IR3 & 17 &  -    & 1 &  -    & 9.3 &  -    & 3 &  -  & 5.2 & 1   \\
IR4 & 13 & 10-16 & 2 & 0.5-4 & 5.6 & 1-8 & 1 & 1-2 & 0.4  & 11 \\
\hline
\label{SEDpar}
\end{tabular}
\end{table}

From the SED we conclude that the three sources are massive protostars. The derived ages are consistent 
with presence of ionized gas as 
detected in the radio continuum emission at 20 cm. It is well known that the formation of an 
ultracompact HII region around a massive protostar requires
timescales about 10$^{5}$ yr \citep{sridh02}. The analysis of the SED also shows that IR2 seems to be 
the most evolved among the three sources. While IR3 and IR4
are protostars that are still accreting material at high rates 
($\dot M_{env} > 5 \times 10^{-5}$ M$_{\odot}$yr$^{-1}$), IR2 probably finished its accretion stage 
($\dot M_{env} \sim 0$). Moreover, from Fig. \ref{RS} it can be seen a shell-like morphology
at 8 $\mu$m towards this source, which shows the border of an incipient 
photodissociation region.
By comparing the age of these YSOs (see Table \ref{SEDpar}) with that of the SNR (about 10$^{5}$ yrs), we                 
conclude that the formation of the YSOs should be approximately coeval with the SN explosion, 
discarding the possibility that the SNR had triggered the star formation in the region.
It it important to note that if the SED is performed using the near distances of 1--3 kpc,
we obtain that the sources should be very young (ages of about $10^{4}$ yrs) with low masses 
(4-6 M$_{\odot}$); stellar sources with these parameters are unable 
to ionize the gas and generate an HII regions complex.

%A likely scenario is that the SNR G18.8+0.3 was the first star exploding in an active region of massive stars.
%An additional hypothesis is that the progenitor star of the SNR might have initiated the formation of massive protostars in the region through
%the action of the powerful stellar wind.

\section{Concluding remarks}

In this work we present the results from  molecular observations (\2 and \3 J=3--2, and 
HCO$^{+}$ J=4--3) obtained with ASTE towards 
a molecular clump belonging to an extended cloud suggested to be interacting with the eastern flank of the SNR G18.8+0.3. 
Our observational study was complemented using IR, submillimeter and radio continuum data from public 
databases. 

The surveyed region contains four radio sources, two of them catalogued as HII regions: G18.751+0.254 
(RS1 in this work) and a compact
one (RS4). The rest of the sources, IR2 and IR3, were identified as new young HII regions.
We found a molecular clump, seen mainly in the \2 J=3--2 line, partially surrounding the HII 
region G18.751+0.254 (RS1) with its densest portion located towards
the north (seen at \3 J=3--2 and HCO$^{+}$ J=4--3). This densest portion coincides with the 
1.1 mm continuum source BGPS 18.763+00.261, and the sources RS2, RS3, and RS4
are embedded in it.

The SNR G18.8+0.3 together with the radio sources RS1 and RS4, and the molecular gas have the same LSR 
velocity of about 20 \k.  
We conclude that all the analyzed objects are located at the same distance of about $14\pm1$ kpc. An important result is that the 
new molecular observations reveal that the surveyed molecular clump is 
not in physical contact with the SNR, suggesting that, either the SNR shock front have not 
yet reached the clump or they are located at different distances. 

We identified the IR counterparts of the radio sources RS2, RS3, and RS4, those sources embedded in 
the densest portion of the molecular clump. From a SED fitting 
we conclude that they are YSOs of ages about 10$^{5}$ yr, which is consistent with the presence of ionized gas
as seen in the radio continuum emission at 20 cm. Their formation should be approximately coeval with 
the SN explosion, discarding triggered star formation by the action of the SNR.

\section*{Acknowledgments}

The ASTE project is driven by Nobeyama Radio Observatory (NRO), a branch
of National Astronomical Observatory of Japan (NAOJ), in collaboration
with University of Chile, and Japanese institutes including University of
Tokyo, Nagoya University, Osaka Prefecture University, Ibaraki University,
Hokkaido University and Joetsu University of Education.
S.P., M.O., and G.D. are members of the {\sl Carrera del 
investigador cient\'\i fico} of CONICET, Argentina.  A.P. is a doctoral fellow of CONICET. 
This work was partially supported by Argentina grants awarded by UBA, CONICET and ANPCYT.
M.R. wishes to acknowledge support from FONDECYT (CHILE) grant No108033.
She is supported by the Chilean {\sl Center for Astrophysics} FONDAP No.
15010003. S.P. and M.O. are grateful to the ASTE staff for the support received during the observations.

%%%%%%%%%%%%%%%%%%%%%%%%%%%%%%%%%%%%%%%%%%%%%%%%%%%%%%%%%%%%%%%%%%%%%
\bibliographystyle{aa}  % A&A format
   %\bibliographystyle{klunamed}     
   % format of references provided by the review (.bst)
\bibliography{biblio}
   % file containing the bibtex references (.bib)
\IfFileExists{\jobname.bbl}{}
{\typeout{}
\typeout{****************************************************}
\typeout{****************************************************}
\typeout{** Please run "bibtex \jobname" to optain}
\typeout{** the bibliography and then re-run LaTeX}
\typeout{** twice to fix the references!}
\typeout{****************************************************}
\typeout{****************************************************}
\typeout{}
}

\label{lastpage}
\end{document}